\documentclass{ws-p9-75x6-50}

\def\leftcontract{\mathop{\hbox{\vrule height0.5pt width6pt \vrule width0.5pt
   height6pt}}}

\def\dualp#1{{}^{\ast_{(\hbox{$\scriptstyle #1$})}} \kern-1pt}
\def\dual{\,{}^\ast{}\kern-1.5pt}
\def\SS{{\cal S}} \def\V{{\cal V}}

\def\undersim#1{\mathop{\vtop{\ialign{##\crcr
     $\hfil\displaystyle{#1}\hfil$\crcr\noalign
     {\kern1pt\nointerlineskip}\hbox{$\hfil{\scriptscriptstyle \sim}\hfil$}\crcr
     \noalign{\kern1pt}}}}}

\def\Lie{\hbox{\it\char'44}} \let\pounds=\Lie

\def\beq{\begin{equation}}
\def\eeq{\end{equation}}
\begin{document}

\title{A CLASSICAL INTERPRETATION OF MAXWELL'S EQUATIONS IN CURVED SPACETIME}

\author{C. Germani}

\address{
School of Computer Science and Mathematics\\
University of Portsmouth, UK\\
and International Center for Relativistic Astrophysics - I.C.R.A.\\
University of Rome ``La Sapienza", I-00185, Roma\\
E-mail: cristiano.germani@port.ac.uk}

\author{D. Bini }

\address{Istituto per Applicazioni della Matematica, CNR, I-80131, Napoli\\
and International Center for Relativistic Astrophysics - I.C.R.A.\\
University of Rome ``La Sapienza", I-00185, Roma\\
E-mail: binid@icra.it}

\author{R.T. Jantzen}

\address{Department of Mathematical Sciences\\
Villanova University, Villanova, PA 19085, USA\\
and
International Center for Relativistic Astrophysics - I.C.R.A.\\
University of Rome ``La Sapienza",I-00185, Roma\\
E-mail: robert.jantzen@villanova.edu
}

\maketitle

\abstracts{We present a new integral formulation of observer-dependent Maxwell's equations in curved spacetime and give a classical interpretation of them.}

\section{Integral formulation of Maxwell's equations}

Let $U$ be an evolution vector field whose 1-parameter
family of diffeomorphisms drags the family ${\cal D}^{(p)}(t)$ into itself, satisfying $\pounds_U \, t = U\leftcontract dt =1$, for a spacetime $p$-form $\Omega$ and $p$-surface integration domain ${\cal D}^{(p)}(t)$ within the hypersurface of (constant) coordinate time $t$. The ``Transport Theorem" \cite{germani} can be written as

\begin{equation} \label{trasporto}
\frac{d}{dt} \int_{{\cal D}^{(p)}(t)}\Omega
= \int_{{\cal D}^{(p)}(t)}\Lie_U\Omega \ .
\end{equation}

Using this theorem for the Maxwell 2-form $F$ and its dual ${}^* F$ when decomposed in terms of electric and magnetic fields $E(u)$, $B(u)$, in a standard way according to the 3+1 splitting associated to an observer family $u$:
\beq
F=u\wedge E(u) +{}^{*(u)}B(u), \qquad ^{*}F=-u\wedge B(u) +{}^{*(u)}E(u),
\eeq
the Stokes' theorem applied to the exterior derivatives $dF$ and $d{}^*F$, and the differential formulation of Maxwell's equation we obtain the Gauss law for the electric field, that for the magnetic field, the Ampere law and the Faraday law respectively:
\begin{eqnarray}
\int_{\partial \V(t)} [{E} (u)+{\nu}(n,u)\times_ u {H} (u)] \cdot dS(u)= 4\pi\int_{\V(t)}  [\rho(u)-{\nu}(n,u)\cdot {J} (u)] \, dV (u)\ ,
\end{eqnarray}
\begin{eqnarray}
\int_{\partial \V(t)} [{H} (u)-{\nu}(n,u)\times_ u {E} (u)] \cdot dS(u)=0,
\end{eqnarray}
\begin{eqnarray}
\label{intfor}
&& -\frac{d}{dt} \int_{\SS(t)} [{H} (u)- {\nu}(n,u)\times_ u {E} (u)] \cdot dS (u) =\nonumber \\
\qquad
&& = \int_{\partial S(t)}
L(u)[E(u)+ {\cal U}\times_u \{ H(u) - \nu(n,u)\times_u E(u) \} ]
\cdot d\ell (u) \ ,
\end{eqnarray}
\begin{eqnarray}
&& -\frac{d}{dt} \int_{\SS(t)}
  [{E} (u)+ {\nu}(n,u)\times_ u {H} (u)] \cdot dS (u)=\nonumber \\
\qquad
&& \int_{\partial S(t)}L(u)[H(u)- {\cal U}\times_u \{ E(u) +
\nu(n,u)\times_u H(u) \} ] \cdot d\ell (u)+ \nonumber \\
\qquad
&& +4\pi\int_{\SS(t)}
L(u)[J(u)-{\cal U}(\rho(u)-\nu(n,u)\cdot_u J (u))] \cdot dS (u) \ ,
\end{eqnarray}
where $n=-Ndt$ ($N$ is a ``lapse" function) is the unit normal to the constant time hypersurfaces, the splitting of which according to the generic observer $u$ identifies the 3-velocity $\nu(n,u)$;
$L(u)=N \gamma (u ,n)^{-1}$; ${\cal U}$ is the 4-velocity of the curve $\partial \SS(t)$ reparametrized by the  proper time of $u$ and projected into the local rest space of $u$.
Let's specialize now equations (\ref{intfor}) to a source free case and for a static integration surface (${\cal U}=0$).
By replacing
$L(u)E(u)=E_u$ and $L(u)H(u)=H_u$
and introducing the following ``constitutive equations" of an equivalent material medium by the fields:
\begin{eqnarray}
B_u=L(u)[H_u-\nu(n,u)\times_u E_u]\ ,\ D_u=L(u)[E_u+\nu(n,u)\times_u H_u]\ .
\end{eqnarray}
one has
\begin{eqnarray}
\int_{\partial \V(t)} D_u \cdot dS(u)&=&0, \qquad
\qquad \int_{\partial \V(t)} B_u \cdot dS(u)=0,\nonumber\\
\int_{\partial S(t)} E_u \cdot d\ell (u)&=&-\frac{d}{dt} \int_{\SS(t)} B_u \cdot dS (u) \ ,\nonumber \\
\int_{\partial S(t)} H_u \cdot d\ell (u)&=&-\frac{d}{dt} \int_{\SS(t)} D_u \cdot dS (u) \ .
\end{eqnarray}
Now one can compare with the integral Maxwell's equations in flat spacetime in presence of matter.
Even if the subject has been widely discussed in the literature\cite{LL}, electric and magnetic fields have always been defined through the coordinates and loss their ``observer-dependent" character. The observer-dependent integral formulation of the Maxwell's equations allows us to ovecome this difficulty.

\end{document}